\documentclass[12.0pt,a4paper]{article}
\usepackage{amsfonts}
\usepackage{latexsym}
\def\Section#1{\section{#1}
\quad \ }

\def \R {{\mathbb R}}
\def \C {{\mathbb C}}
\def \Z {{\mathbb Z}}

\def \lb {\left(}
\def \rb {\right)}
\def \Lb {\left[}
\def \Rb {\right]}
\def \LB {\left\{}
\def \RB {\right\}}
\def \l. {\left.}
\def \r. {\right.}

\begin{document}

\title{
{\Large \bf Dimensional Reduction of Dirac Operator} }
\author{
Petko A. Nikolov {\normalsize ${\!}{\,}^{{\,}^{a)}}$}\footnote{
pnikolov@phys.uni-sofia.bg } \quad and \quad Gergana R. Ruseva
{\normalsize ${\!}{\,}^{{\,}^{b)}}$}\footnote{
gergana@inrne.bas.bg }
\\ \\
{\normalsize \small \( \begin{array}{l} {\!\!\!\!\!}{\,}^{a)} \,
\mathrm{ Department \ of \ Theoretical \ Physics, \ Sofia \
University } \\ \, \mathrm{ James \ Bourchier \ 5, \ BG \!\! -
\!\! 1164 \ Sofia, \ Bulgaria }
\\ \\ {\!\!\!\!\!}{\,}^{b)} \,
\mathrm{ Institute \ for \ Nuclear \ Research \ and \ Nuclear \
Energy, } \\ \, \mathrm{ Tsarigradsko \ Chaussee \ 72, \ BG \!\! -
\!\! 1784 \ Sofia, \ Bulgaria }
\end{array} \) }}
\maketitle
\begin{abstract}
We construct an explicit example of dimensional reduction of the
free massless Dirac operator with an internal $\mathrm{SU}(3)$
symmetry, defined on a 12-dimensional manifold that is the total
space of a principal $\mathrm{SU}(3)$-bundle over a
four-dimensional (nonflat) pseudo-Riemannian manifold. Upon
dimensional reduction the free twelve-dimensional Dirac equation
is transformed into a rather nontrivial four-dimensional one: a
pair of massive Lorentz spinor $\mathrm{SU}(3)$-octets interacting
with an $\mathrm{SU}(3)$-gauge field with a source term depending
on the curvature tensor of the gauge field. The $\mathrm{SU}(3)$
group is complicated enough to illustrate features of the general
case. It should not be confused with the color $\mathrm{SU}(3)$ of
quantum chromodynamics where the fundamental spinors, the quark
fields, are $\mathrm{SU}(3)$ triplets rather than octets.
\end{abstract}

\Section{Introduction} It is well known that when we look for a
solution with some symmetry, we can reduce the number of variables
and thus simplify the problem of solving differential equations.
The Schwarzschild solution of the nonlinear Hilbert-Einstein
equation is a typical example. A point of view, different from
this calculational aspect of symmetry, is essential for the so
called "Kaluza-Klein approach." It is observed in the pioneer work
of Kaluza (1921, English translation in \cite{Lee}) that there is
one-to-one correspondence between the $\mathrm{U}(1)$-invariant
metrics on a five-dimensional manifold and the triples \{metric on
four-dimensional manifold, linear connection with structure group
$\mathrm{U}(1)$ (electromagnetic potential), scalar field\}. The
scalar curvature of five-dimensional $\mathrm{U}(1)$-invariant
metric is equivalent to the Einstein-Maxwell action for the
mentioned fields. This action describes the really observed
interaction between gravity and electromagnetic field. This
demonstrates the general idea: We consider a ``simple'' field and
``simple'' equations but in a ``multidimensional'' universe.
Imposing some symmetry conditions, after dimensional reduction we
obtain a set of fields with different nature involved in
complicated differential equations. Our hope is that the fields
and differential equations, obtained in this way, may describe a
real process, and that this investigation may be a step to the
unification of different interactions in nature. The natural
generalizations of the Kaluza- Klein ansatz are considered in the
literature: the group of symmetry $G$ is arbitrary, the group $G$
acts on a manifold as on a total space of a principal bundle, and
the group $G$ acts on a manifold with one type orbits. See, for
example \cite{CJ}.

  In this paper the starting point is the free Dirac operator with
  an $\mathrm{SU}(3)$ symmetry defined on a twelve-dimensional
  Minkowski space that is interpreted as an $\mathrm{SU}(3)$
  principal bundle over four-dimensional Minkowski space.
  Should we interpret the outcome in physical terms
  we should relate the structure group with the
  "flavor $\mathrm{SU}(3)$" of the quark model, identifying the
  resulting $\mathrm{SU}(3)$-octets Dirac particles with observed baryons.
  Such an interpretation would again be a nonstandard one however since,
  unlike the flavor $\mathrm{SU}(3)$ of the standard model, our structure
  group appears as a local gauge group in four-space-time. We prefer, in fact,
  to view the present paper as a mathematical model
  illustrating some surprising features of dimensional reduction.

  Our purpose is to consider the simplest possible case because
  then the arising structures after
dimensional reduction are imperative. The initial manifold,
denoted by $E$ in the text, is the twelve-dimensional total space
of a principal $\mathrm{SU}(3)$ bundle, which admits a real spinor
bundle with standard fibre $\R^{64}$. In the real case, the spinor
connection is uniquely defined if it is compatible with the
Levi-Civita connection of the metric on $E$. For physical reasons
we consider a complex spinor bundle, a complexification of the
real-valued one. The spinor connection is also considered as a
complexification of the real one. Thus we avoid the necessity to
fix a connection with structure group $\mathrm{U}(1)$. Further,
when we fix the $\mathrm{SU}(3)$-action on spinor fields we choose
the trivial lifting. And thus we avoid some additional terms in
the reduced Dirac operator. Also the scalar field in the
Kaluza-Klein ansatz is taken to be constant - the Killing metric
in the Lie algebra of $\mathrm{SU}(3)$. In this way, in the
reduced Dirac operator there are only structures whose presence is
necessary. We also point out the steps in which, imposing the
symmetry, the new structures arise (the gauge field with structure
group $G=\mathrm{SU}(3)$, its curvature tensor, the Clifford
algebra for a four-dimensional manifold, the four-dimensional
Dirac operator, the spinor octets, the mass term etc.).

We choose the group of symmetry to be $\mathrm{SU}(3)$ acting
freely on the 12D manifold because of its connection to the
standard model and because we wanted the arising after the
reduction gauge field to have structure group $\mathrm{SU}(3)$. In
the same way one can obtain the dimensional reduction of Dirac
operator when the symmetry group is an arbitrary connected Lie
group acting freely on the multidimensional manifold.
\\

The article is organized as follows.

In \textbf{Section 2} the necessary constructions from
differential geometry and the algebraic origin of the Kaluza-Klein
ansatz are presented. We give the coordinate expression of the
Levi-Civita connection for the metric in nonholonomic basis. These
formulas are applied to the canonical basis of the
$\mathrm{SU}(3)$-invariant metric on $E$. This basis determines a
horizontal subbundle $T^h(E)\hookrightarrow T(E)$. The subbundle
$T^h(E)$ is invariant under the action of $\mathrm{SU}(3)$ and
defines a linear connection (gauge field with structure group
$\mathrm{SU}(3)$). The components of the Levi-Civita connection
for the $\mathrm{SU}(3)$-invariant metric are calculated and they
contain components of the gauge field and its stress tensor
(eq.(\ref{16})).

In \textbf{Section 3} the Dirac operator for the
$\mathrm{SU}(3)$-invariant metric (the Kaluza-Klein ansatz) is
considered. The crucial moment here is that the sum
$T(E)=T^h(E)\oplus T^v(E)$ is orthogonal with respect to the
$\mathrm{SU}(3)$-invariant metric. According to the classifying
theorem for Clifford algebras, the Clifford algebra
$\mathrm{Cl}(T_z(E), g(x))\approx M_{64}(\R)$ is realized as a
tensor product of the Clifford algebras of $T_z^h(E)$ and
$T_z^v(E)$. So the standard fibre $\C^{64}$ of the spinor bundle
on $E$ takes the structure of tensor product $\C^4\otimes\C^{16}$.

In \textbf{Section 4} we give the dimensional reduction of the
Dirac operator for the $\mathrm{SU}(3)$-invariant metric. We
introduce an action of $\mathrm{SU}(3)$ on the spinor bundle,
compatible with the action of $\mathrm{SU}(3)$ on $T(E)$. This
condition of compatibility does not fix uniquely the action of
$\mathrm{SU}(3)$ on the spinors. So we choose, as we mentioned
above, the simplest case in which the lifting of the
$\mathrm{SU}(3)$ action on $E$ to the total space of the spinor
bundle is trivial in the canonical basis.

In \textbf{Section 5} we list the steps in the procedure of
dimensional reduction where the new structures presented in the
reduced Dirac operator arise.

\Section{Basic constructions and notations} Let $E$ be a smooth
manifold, $g$ a metric on $E$ (with arbitrary signature), and
$\nabla$ the corresponding Levi-Civita connection. Let
$\{\mathbf{h}_{\mu}\}$ be a (local) nonholonomic basis of $T(E)$
and $\{\mathbf{h}^{\mu}\}$ the corresponding dual basis on
$T^*(E)$. In this basis we have the following notation:
$$\begin{array}{ll}
 \nabla(\mathbf{h}_{\beta})=\Gamma_{\alpha\beta}^{\,\rho}\mathbf{h}^{\alpha}\otimes \!
\mathbf{h}_{\,\rho}\,\,, \;\;&
\nabla_{\!\mathbf{h}_\alpha}(\mathbf{h}_{\beta})=
\nabla_{\alpha}(\mathbf{h}_{\beta})= \Gamma_{\alpha\beta}^{\,\rho}\mathbf{h}_\rho\,,\\
\Lb\mathbf{h}_{\alpha},\mathbf{h}_{\beta}\Rb =
C_{\alpha\beta}^{\rho}\,\mathbf{h}_{\rho}\,,
\;\;& g(\mathbf{h}_{\alpha}, \mathbf{h}_{\beta})=g_{\alpha\beta}\,,\\
g(\nabla_{\alpha}(\mathbf{h}_{\beta}),
\mathbf{h}_{\gamma})=\Gamma_{\alpha\beta}^{\rho}g_{\rho\gamma}=
\Gamma _{\alpha\beta\gamma}\, , \;\;&
g([\mathbf{h}_{\alpha},\mathbf{h}_{\beta}], \mathbf{h}_{\gamma})=
C_{\alpha\beta}^{\rho}g_{\rho\gamma}=
C_{\alpha\beta\gamma}\, .\\
\end{array}$$
The condition $\nabla(g)=0$ and the requirement for the vanishing
of the torsion reads:
\begin{equation}
\begin{array}{lll}
  \mathbf{h}_{\rho}(g_{\mu\nu})=g(\nabla_{\rho}(\mathbf{h}_{\mu}), \mathbf{h}_{\nu})+g(\mathbf{h}_{\mu},
\nabla_{\rho}(\mathbf{h}_{\nu})) & \Rightarrow &
\Gamma_{\rho\mu\nu}+\Gamma_{\rho\nu\mu}=
\mathbf{h}_{\rho}(g_{\mu\nu})\\
 \nabla_{\mu}(\mathbf{h}_{\nu})-\nabla_{\nu}(\mathbf{h}_{\mu})=[\mathbf{h}_{\mu}\,,\,\mathbf{h}_{\nu}]
& \Rightarrow & \Gamma_{\mu\nu\rho}-\Gamma_{\nu\mu\rho}=C_{\mu\nu\rho}\, , \\
\end{array}
\end{equation}
and from here it follows that
\begin{equation}\label{2}
 2\Gamma_{\alpha\beta\gamma}=C_{\alpha\beta\gamma}+C_{\gamma\beta\alpha}+C_{\alpha\gamma\beta}+
\mathbf{h}_{\alpha}(g_{\beta\gamma})+\mathbf{h}_{\beta}(g_{\alpha\gamma})-
\mathbf{h}_{\gamma}(g_{\alpha\beta})\, .
\end{equation}
We follow the classical construction of the generalized
Kaluza-Klein ansatz. The point structure of the ansatz is the
description of the metric on the vector space $L$, which is the
middle term in the short exact sequence:
\begin{equation}\label{3}
 0\quad \longrightarrow \quad L_0 \quad \mathop{\longrightarrow} \limits^i \quad L \quad
\mathop{\longrightarrow}\limits^j \quad L_1 \quad \longrightarrow
\quad 0  \ .
\end{equation}
We realize this in coordinates by choosing a basis
$\LB\mathbf{h}\RB=\LB \mathbf{f},\mathbf{e} \RB $ in $L \,$; \quad
$\mathbf{h}_{\mu}= \mathbf{f}_{\mu},\
\mu=1,2,...,m=\mathrm{dim}(L_1),\ \ \mathbf{h}_{k}= \mathbf{e}_k,
\ k=m+1,...,m+n, n=\mathrm{dim}(L_0)$. The vector space
$L_0=\mathrm{span}(\mathbf{e}_1,...,\mathbf{e}_n)$, and the vector
space $L_1$ is identified with
$\mathrm{span}(\mathbf{f}_1,...,\mathbf{f}_m)$ and
$i(\mathbf{e}_k)=\mathbf{e}_k$; $j(\mathbf{e}_k)=\mathbf{0}$,
$j(\mathbf{f}_\mu)=\mathbf{f}_\mu$. Every splitting of the exact
sequence (\ref{3}) is given by a linear map $S:L_1\, \rightarrow
\, L$ with the property $j\circ S=\mathbf{1}$, i.e., is given by
defining the vectors:
\begin{equation}\label{4}
\widehat{\mathbf{f}}_\mu=S(\mathbf{f}_\mu)=\mathbf{f}_\mu-A_\mu^k\mathbf{e}_k
\, .
\end{equation}
In these formulas we have summation over repeated indices. Here
the matrix ${A_\mu^k}$ is arbitrary. Every metric $g_{_L}$ on $L$,
for which the restriction on $i(L_0)$ is nondegenerate is uniquely
determinate by the conditions:
\begin{equation}\label{5}
    \begin{array}{ll}
      g_{_L}(\widehat{\mathbf{f}}_\mu, \widehat{\mathbf{f}}_\nu) = g_{\mu\nu}\, , \\
      g_{_L}(\widehat{\mathbf{f}}_\mu, \mathbf{e}_k) = 0 \, ,\\
      g_{_L}(\mathbf{e}_k,\mathbf{e}_l)=g_{0 \,kl}  \, , \\
    \end{array}
\end{equation}
where $g_{0 \,kl}$ and $g_{\mu\nu}$ are metrics on $L_0$ and
$L_1$. In this manner we have one-to-one correspondence  between
the metrics on $L$, nondegenerated on $L_0$ and the triples
\{metric on $L_1$, metric on $L_0$, splitting of (\ref{3})\}.
\\
In the basis $\{{\mathbf{f}_\mu,\mathbf{e}_k}\}$ the metric,
defined by the equations (\ref{5}) has components:
\begin{equation}\label{6}
\{g_{_L}\}=
    \left(%
\begin{array}{cc}
  g_{\mu\nu}+A_\mu^{\,\,i}\,A_\nu^{\,\,j}\,g_{0\, ij} & A_\mu^{\,\,i}\, g_{0\, il} \\
  g_{0\, ki}\,A_\nu^i & g_{0\,kl}  \\
\end{array}
\right)\, .
\end{equation}
The above construction is the algebraic origin of the Kaluza-Klein
ansatz. In the case of the general Kaluza-Klein ansatz this
construction arises in the tangent space of each point of the
manifold where the group of symmetry acts. More precisely, let
$(E,p,M)$ be a principal bundle with structure group
$G=\mathrm{SU}(3)$. We assume for simplicity that the principal
bundle is trivial and the manifold $M$ is isomorphic to $\R^4$ as
a topological manifold. We take a global trivialization $E=M\times
\mathrm{SU}(3)$ and the right group action is
$R_\mathbf{g}(\mathbf{x},\mathbf{z})=(\mathbf{x},\mathbf{zg})$
where $(\mathbf{x},\mathbf{z})\in M\times \mathrm{SU}(3)$ and
$\mathbf{g} \in \mathrm{SU}(3)$. We also fix coordinates on the
total space $E$, $(x^\mu,z^k)=(x,z)$, $\mu=1,2,3,4 $,
$k=5,...,12$, $p(x^\mu, z^k)=(x^\mu)$. Because of the assumed
triviality of $M$ the coordinates $x^\mu$ are global. Let
$T^v(E)\hookrightarrow T(E) $ be the vertical subbundle,
$\mathbf{f}_\mu$ nonholonomic basis of $T(M)$, $\mathbf{e}_k$ -
the fundamental fields on $E$ corresponding to a basis
$\widehat{\mathbf{e}}_5,...,\widehat{\mathbf{e}}_{12}$ of the Lie
algebra ${su(3)}$. The fields $\mathbf{f}_\mu,\mathbf{e}_k$ form a
nonholonomic basis on $TE=T(M \times \mathrm{SU}(3))$, and have
the form
\begin{equation}\label{7}
\mathbf{f}_\mu=f^{ \, \nu} _\mu (x)\frac{\partial}{\partial
x^\nu}\, \, , \quad \mathbf{e}_k=e_k^{ \, \, l}(z)
\frac{\partial}{\partial z^l} \; .
\end{equation}
The natural exact sequence
\begin{equation}\label{8}
    0 \quad \longrightarrow \quad T^v(E) \quad \longrightarrow
\quad T(E) \quad \longrightarrow \quad p^\ast (T(M)) \quad
\longrightarrow \quad 0
\end{equation}
realizes the exact sequence (\ref{3}) at the tangent space of each
point of the manifold $E$. Here $p^\ast (T(M))$ is the pull back
of the tangent bundle of $M$ (see \cite{Gd}). Each metric $g_{_E}$
on $E$ can be written in the form:
\begin{equation}\label{9}
 \{g_{_E}(x,z)\}=
    \left(
\begin{array}{cc}
  g_{\mu\nu}(x,z)+A_\mu^{\,\,i}(x,z)\,A_\nu^{\,\,j}(x,z)\,g_{0\, ij}(x,z) & A_\mu^{\,\,i}(x,z)\, g_{0\, il}(x,z) \\
  g_{0\, ki}(x,z)\,A_\nu^i(x,z) & g_{0\,kl}(x,z)  \\
\end{array}
\right)\,,
\end{equation}
where
$g_{\mu\nu}(x,z)=g_{_E}(x,z)(\mathbf{f}_\mu(x),\mathbf{f}_\nu(x))$,
\quad $
g_{0\,kl}(x,z)=g_{_E}(x,z)(\mathbf{e}_k(z),\mathbf{e}_l(z))$. The
vector fields
$\widehat{\mathbf{f}}_\mu(x,z)=\mathbf{f}_\mu(x)-A_\mu^{\,\,k}(x,z)\mathbf{e}_k(z)
$ span a horizontal subbundle $T^h(E)\hookrightarrow T(E)$
orthogonal to $T^v(E)$ with respect to the metric (\ref{9}). The
ansatz (\ref{9}) is convenient to describe the metrics on $E$
invariant under the action of the group $G=\mathrm{SU}(3)$. The
invariant metrics $g$ on $E$ have the form
\begin{equation}\label{10}
 \{g_{_E}(x,z)\}=
    \left(
\begin{array}{cc}
  g_{\mu\nu}(x)+A_\mu^{\,\,i}(x)\,A_\nu^{\,\,j}(x)\,g_{0\, ij}(x) & A_\mu^{\,\,i}(x)\, g_{0\, il}(x) \\
  g_{0\, ki}(x)\,A_\nu^i(x) & g_{0\,kl}(x)  \\
\end{array}
\right)\, .
\end{equation}
\quad This is the Kaluza-Klein ansatz in our case. In this formula
$g_{\mu\nu}(x)= g(\mathbf{f}_\mu,\mathbf{f}_\nu)(x)$ is an
arbitrary metric on $M$. Because $\mathbf{f}_{\mu}$ is a
nonholonomic basis on $M$, without loss of generality we can think
that $g_{\mu\nu}$ is in canonical form, i.e., $\mathbf{f}_\mu$ are
tetrada. This will be used in our calculation later. $g_{0 \,
kl}(x)$ at each point $x\in M$ is invariant metric on the Lie
algebra $su(3)$, i.e., $g_0$ is a field defined on $M$ taking
values in the set of invariant metrics on the Lie algebra $su(3)$.
The vector fields
$\widehat{\mathbf{f}}_\mu(x,z)=\mathbf{f}_\mu(x)-A_\mu^{\,\,k}(x)\mathbf{e}_k(z)
$ span orthogonal horizontal subbundle $T^h(E)$ which is invariant
under the action of the structure group of the principal bundle
$G=\mathrm{SU}(3)$. So $A_\mu^k (x)$ define a linear connection in
the principal bundle with a structure group $G=\mathrm{SU}(3)$. A
classical result is that there is one-to-one correspondence
between the $G$-invariant metrics on $E$ and the triples \{metric
on $M$, linear connection with values in the Lie algebra of $G$,
``scalar field''\}. In the basis
$\{\widehat{\mathbf{f}}_\mu,\mathbf{e}_k\}$
 for the metric (\ref{10}) we have
\begin{equation}\label{11}
    \begin{array}{ll}
      g_{_E}(\widehat{\mathbf{f}}_\mu(x), \widehat{\mathbf{f}}_\nu(x))=
g(\mathbf{f}_\mu(x), \mathbf{f}_\nu(x))=g_{\mu\nu}(x)\, ,\\
      g_{_E}(\widehat{\mathbf{f}}_\mu(x), \mathbf{e}_k(x))=0 \, ,\\
      g_{_E}(\mathbf{e}_k(z), \mathbf{e}_l(z))=g_{0 \, kl}(x) \, . \\
    \end{array}
\end{equation}
The next step is to construct the Dirac operator on $E$
corresponding to the metric (\ref{10}). To calculate the
Levi-Civita connection (\ref{2}) of the metric (\ref{10}) we have
to introduce the commutator coefficients for the basis
$\{\mathbf{f}_\mu , \mathbf{e}_k\}$:
\begin{equation}\label{12}
  \begin{array}{ll}
      \Lb \mathbf{f}_\mu ,\mathbf{f}_\nu \Rb  (x) = C_{\mu \nu}^{\,\,\,\, \rho }(x)\mathbf{f}_\rho (x) \, \\
     \Lb \mathbf{f}_\mu ,\mathbf{e}_k \Rb =
   \Lb f_\mu ^{\,\, \nu } (x)\frac{\partial}{\partial x^\nu },
     e_k^{\,\,l} (z)\frac{\partial}{\partial z^l} \Rb =0 \, ,\\
    \Lb \mathbf{e}_k,\mathbf{e}_l \Rb (z)=t_{kl}^{\,\,m}\mathbf{e}_m (z) \, . \\
    \end{array}
\end{equation}
Here $C_{\mu\nu}^{\,\,\rho}(x)$ are determined by the choice of
the nonholonomic basis ${\mathbf{f}_\mu}$ in $T(M)$,
$t_{kl}^{\,\,m}$ are the structure constants of the Lie algebra
$su(3)$ for the basis
$\widehat{\mathbf{e}}_5,...,\widehat{\mathbf{e}}_{12}$. The
nonholonomic basis $\{{\widehat{\mathbf{f}}_\mu,\mathbf{e}_k}\}$,
because of (\ref{11}), is convenient for the construction of the
Dirac operator. By means of (\ref{12}) and (\ref{4}) we calculate
\begin{equation}\label{13}
    \begin{array}{ll}
       \Lb \widehat{\mathbf{f} }_\mu ,\widehat{\mathbf{f} }_\nu \Rb (x,z)=
C_{\mu \nu}^{\,\,\,\,\rho}(x)\widehat{\mathbf{f} }_\rho (x)-F_{\mu \nu}^{\,\,\,k}(x)\mathbf{e}_k(z)\, ,\\
       \Lb \widehat{\mathbf{f}}_\mu, \mathbf{e}_k \Rb (x,z)=-A_{\mu}^{\,l}(x)t_{lk}^{\,\,m}\mathbf{e}_m(z)\, , \\
       \Lb \mathbf{e}_k,\mathbf{e}_l \Rb (z)=t_{kl}^{\,\,m}\mathbf{e}_m \, .  \\
    \end{array}
\end{equation}
Here
\begin{equation}\label{14}
   F_{\mu\nu}^{\; \,m}=\mathbf{f}_\mu (A_\nu ^m)-\mathbf{f}_\nu (A_\mu ^m) +
 t_{\mu\nu}^{\;\; \rho} A_\rho ^m - C_{\mu\nu} ^{\; \; \rho} A_\rho ^m
\end{equation}
is the curvature tensor of the linear connection, determined by
the one-form $A_\mu=A_\mu ^m a_m$. Then the coefficients
$C_{\alpha \beta \gamma }$ in (\ref{2}) for the basis $\{
\widehat{\mathbf{f}}_\mu ,\mathbf{e}_k\}$ are
\begin{equation}\label{15}
    \begin{array}{ll}
      C_{\mu \nu \rho}=g(\Lb \mathbf{f}_\mu,\mathbf{f}_\nu \Rb , \mathbf{f}_\rho )=
C_{\mu\nu}^{\;\;\sigma}\, g_{\sigma\rho}\, , \qquad
& C_{\mu\nu k}=-g_{0\, kl}F_{\mu\nu}^{\;\;l}=-F_{\mu\nu k} \\
      C_{\mu k \nu}=-C_{k \mu \nu}=0 \, , & C_{\mu k l}=-A_\mu^{\,k}\, t_{mkl}=-C_{k \mu l} \\
      C_{kl\mu}=0 \, ,& C_{klm}=t_{klm}\, ,  \\
    \end{array}
\end{equation}
where $t_{mkl}=g_{0\, li}t_{mk}^{\;\; i}$. From (\ref{2}) we
obtain the components of Levi-Civita connection in the basis
$\{\widehat{\mathbf{f}}_\mu, \mathbf{e}_k \}$:
\begin{equation}\label{16}
\begin{array}{ll}
  \Gamma ^E_{\mu\nu\rho}=\Gamma _{\mu\nu\rho}\, ,
& \Gamma ^E_{\mu\nu k}= -\frac{1}{2}F_{\mu \nu k}\\
  \Gamma ^E_{\mu k \nu}= \frac{1}{2}F_{\mu \nu k}\, ,
& \Gamma ^E_{\mu kl}=\frac{1}{2} A_\mu ^m (t_{mlk}-t_{mkl})+ \frac{1}{2} \mathbf{f}_\mu(g_{0\;kl})\\
  \Gamma ^E_{k \mu\nu}= \frac{1}{2}F_{\mu\nu k}\, ,
& \Gamma ^E_{k \mu l}= \frac{1}{2}A_\mu^m(t_{mkl}+t_{mlk})+ \frac{1}{2}\mathbf{f}_\mu(g_{0\;kl}) \\
  \Gamma ^E_{kl \mu}=-\frac{1}{2}A_\mu ^m (t_{mkl}+t_{mlk})+\frac{1}{2} \mathbf{f}_\mu(g_{0\;
  kl})\, ,
& \Gamma ^E_{klm}=\frac{1}{2}(t_{klm}+t_{mlk}+t_{mkl}) \, . \\
\end{array}
\end{equation}
In these formulas $\Gamma _{\mu\nu}^\rho$ are the components of
the Levi-Civita connection in the basis $\{\mathbf{f}_\mu\}$ for
the metric $g$ on $M$.

\Section{Dirac operator for the Kaluza-Klein metric} To describe
the Dirac operator for the Kaluza-Klein metric (\ref{10}) we need
some preliminary constructions. Let $L$ be a real vector space and
$\eta$ - a metric on $L$ of type $(p,q)$;
$\eta=\mathrm{diag}(-1,...,-1,1,...,1),\,\,$ \{number of
$(-1)\}=p,\,\,$ \{number of $(1)\}=q, \,\, p+q=\mathrm{dim}(L)$.
We denote by $\vartheta$ the canonical embedding
\begin{equation}\label{17}
    \vartheta:L \; \longrightarrow \; \mathrm{Cl}(\eta)
\end{equation}
of the vector space $L$ into the corresponding Clifford algebra
\cite{ABS}, \cite{LM}. $\vartheta(\mathbf{x})^2=\eta(\mathbf{x} ,
\mathbf{x})\mathbf{1}, \;\mathbf{x} \in L $, where $\mathbf{1}$ is
the unit of the algebra $\mathrm{Cl}(\eta)\equiv \mathrm{Cl}^{p,q}
$. If $\mathbf{a}_1,...,\mathbf{a}_n$ is an arbitrary basis of
$L$, $\vartheta (\mathbf{a}_i)=\gamma_i , \; \gamma_i
\gamma_j-\gamma_j \gamma_i=2\eta_{ij}\mathbf{1} , \;
\eta_{ij}=\eta(\mathbf{a}_i, \mathbf{a}_j)$. If
$\mathbf{a}_1,...,\mathbf{a}_n$ is oriented orthonormal basis, the
volume element $\omega=\gamma_1...\gamma_n$ is uniquely determined
and $\omega ^2=\pm\mathbf{1}$. We will assume that $L$ has fixed
orientation. The symmetry of the metric on $L$ gives rise to some
structures on the spinor bundle on $L$. In order to describe them
we need some facts for the classification  of the Clifford
algebras.

The first step in the classification and realizations of the
Clifford algebras for arbitrary metric is the following statement
(\cite{ABS}, \cite{LM}): If $(L_1\oplus L_2, \eta_1\oplus \eta_2)$
is an orthogonal direct sum of metric vector spaces $(L_1,
\eta_1)$ and $(L_2, \eta_2)$ then $\mathrm{Cl}(\eta_1 \oplus \,
\eta_2) = \mathrm{Cl}(\eta_1)\, \widehat{\otimes}\,\,
\mathrm{Cl}(\eta_2)$, where $\, \widehat{\otimes} \;$ is the
$\Z_2$-graded tensor product of the naturally $\Z_2$-graded
Clifford algebras. In some exceptional cases the  $\Z_2$-graded
tensor product may be replaced with the usual tensor product. In
our example is realized one of these exceptional cases.

Let $\mathrm{dim}(L)=p+q=2k$ be even. We say that $
\mathrm{Cl}(\eta)>0$ or $\mathrm{Cl}(\eta)<0$ if $\omega^2=+ 1$ or
$\omega^2=-1$. Let $(L_1,\eta_1)$ and $(L_2, \eta_2)$ be vector
spaces with metrics $\eta_1$ and $\eta_2$, and $(L_1\oplus L_2,
\eta_1\oplus\eta_2)$ be an orthogonal direct sum. Then (see
\cite{ABS}, \cite{LM})
\begin{equation}\label{18}
    \begin{array}{l}
      \mathrm{Cl}(\eta_1)>0 \;\; \Longrightarrow \mathrm{Cl}(\eta_1 \oplus \eta_2)=
      \mathrm{Cl}(\eta_1) \otimes \mathrm{Cl}(\eta_2) \, ,\\
      \mathrm{Cl}(\eta_1)<0 \;\; \Longrightarrow \mathrm{Cl}(\eta_1 \oplus \eta_2)=
      \mathrm{Cl}(\eta_1) \otimes \mathrm{Cl}(-\eta_2) \, .\\
    \end{array}
\end{equation}
Let $\{\mathbf{a}_ {_{\!1}i}\}, \{ \mathbf{a}_ {_{\!2}j}\}$ be
bases in $L_1$ and $L_2$. The isomorphisms in (\ref{18}) are given
by
\begin{equation}\label{19}
    \begin{array}{l}
      \gamma_ {_{\!1}i} \otimes \mathbf{1}_2 \;\longmapsto \;
        \gamma_i=\vartheta( \mathbf{a}_ {_{\!1}i}, \mathbf{0}) , \; i=1,2,...,n_1=\mathrm{dim}(L_1)\\
      \omega_1 \otimes \gamma_ {_{\!2}j}, \longmapsto \;
\gamma_{n_1+j}=\vartheta(\mathbf{0}, \mathbf{a}_ {_{\!2}j})
      , \; j=1,2,...,n_2=\mathrm{dim}(L_2) \; . \;  \\
    \end{array}
\end{equation}
These isomorphisms give the classification of the Clifford
algebras.
\\
\quad For physical reasons we will consider complex spinor fields
and we will need a complexification of the Clifford algebra:
\begin{equation}\label{20}
    \mathrm{Cl}^{p,q} \otimes \C = \mathrm{Cl}(L \otimes \C, \eta)=\mathrm{Cl}^n \, ,\quad \;\; \; \; n=p+q \, .
\end{equation}
It is known (\cite{ABS}, \cite{LM}) that $ \mathrm{Cl}^{2k}
\approx M_{2^k}(\C), \;\; \mathrm{Cl}^{2k+1} \approx M_{2^k}(\C)
\oplus M_{2^k}(\C) $. In our case the Clifford algebra
$\mathrm{Cl}^{12}$ is even and thus $ \mathrm{Cl}^{12} \approx
M_{64}(\C)$ has only one simple module. We have the following
isomorphisms:
\begin{equation}\label{21}
    \begin{array}{ll}
      \mathrm{Cl}^{1,3}  \approx M_4(\R)\;, & \mathrm{Cl}^{1,3}<0 \\
      \mathrm{Cl}^{0,8}=\mathrm{Cl}^{8,0}  \approx M_{16}(\R)\; , &
      \mathrm{Cl}^{0,8}>0\, ,\; \quad \mathrm{Cl}^{8,0}>0 \; . \\
    \end{array}
\end{equation}
So we have
\begin{equation}\label{22}
    \mathrm{Cl}^{1,11}=\mathrm{Cl}^{1,3} \otimes \mathrm{Cl}^{8,0}  \approx M_4(\R) \otimes M_{16} (\R) \; .\;
\end{equation}

Let $(E,g)$ be an oriented even dimensional manifold with metric
$g$, $\mathrm{sign}(g)=(p,q)$. This means that the tangent  bundle
over $E$ has a cocycle $\psi_{\alpha \beta }(x) \in \mathrm{Aut}
(\R^{2k},\eta)=O(p,q)$. An element $A \in \mathrm{Aut} (\R^{2k},
\eta)$ uniquely determines an element $\widetilde{A}\in
\mathrm{Aut} (\mathrm{Cl} ^{p,q})$. So $\widetilde{\psi}(x) \in
\mathrm{Aut} (\mathrm{Cl}^{p,q})$ is a cocycle which defines the
Clifford bundle
 $\mathrm{Cl}(TE)$ over $E$. In this bundle the fiber $\mathrm{Cl}(TE)_x$
 is the Clifford algebra for the vector space
$(T_x(E), g(x))$. The standard fiber of the complex Clifford
bundle $\mathrm{Cl}^\C (TE)$ is $\mathrm{Cl}^{2k}=M_{2^k}(\C)$. A
$\mathrm{spin}^\C$ structure on $E$ is equivalent to a bundle
$\zeta$ of simple complex modules over the Clifford bundle
$\mathrm{Cl}^\C(TE) \,$. We will point out some details in the
construction of the complexified bundle $\zeta$ because they are
important in our later study on the structures arising in $\zeta$.
Let $\widehat{L}=\vartheta(L) \hookrightarrow \mathrm{Cl}(\eta)$
be the image of the linear space $L$ in the Clifford algebra.
$\vartheta\;:\;L \longrightarrow \;\widehat{L}$ is an isomorphism
and we will identify $L$ and $\widehat{L}$ by means of
$\vartheta$. The Clifford group $\mathcal{F}$ is defined by
\begin{equation}\label{23}
    \mathcal {F}^{p,q}=\{c\in \mathrm{Cl}^\ast (p,q)\ \mid \ c\;\widehat{L}\; c^{-1}\subset L\}\;,
\end{equation}
 where $\mathrm{Cl}^\ast $ is the set of invertible elements. In the complexified case we have similarly
\begin{equation}\label{24}
   \mathcal{F}^{2k}= \{c \in \mathrm{Cl} ^{\ast 2k} \ \mid\  c\;\widehat{L}\;c^{-1}\subset L \}\, .
\end{equation}
The linear map $j(c)\;:\widehat{L}\longrightarrow \widehat{L} \;
;$ $\mathbf{u}\; \longrightarrow
\;j(c)(\mathbf{u})=c\;\mathbf{u}\;c^{-1}$ is orthogonal and
$j\;:\mathcal{F} \; \longrightarrow\; O(p,q)$ is surjective and
gives the exact sequence
\begin{equation}\label{25}
    1 \;\longrightarrow \; \C^\ast \;\longrightarrow \; \mathcal{F}^{2k}\mathop{\longrightarrow}
\limits^{j^\C} \;O(p,q)\; \longrightarrow \; \mathbf{1} \, .
\end{equation}
And in the real case we have
\begin{equation}\label{26}
    1 \;\longrightarrow \; \R^\ast \;\longrightarrow \; \mathcal{F}^{p,q}\mathop{\longrightarrow} \limits^j
\;O(p,q)\; \longrightarrow \; \mathbf{1} \, .
\end{equation}

A  $\mathrm{spin}^\C$ bundle $\zeta$ is a complex vector bundle on
$E$ and each fiber $\zeta_{(\mathbf{x},\mathbf{z})}, \;
(\mathbf{x},\mathbf{z})\in E=M \otimes G$, is a simple complex
module over the algebra $\mathrm{Cl}(T_{(x,z)}E)$. Let
$\psi_{\alpha \beta }(x) \in \mathrm{Aut} (\R^{2k},\eta)$ be a
cocycle of $T(E)$ and $\widetilde{\psi}_{\alpha \beta}(x) \in
\mathrm{Aut}(\mathrm{Cl}^{2k})$ be the cocycle of $\mathrm{Cl}^\C
(TE)$. Because the standard fiber of $\mathrm{Cl}(TE)$ is
$\mathrm{Cl}^{2k}\approx M_{2^k}(\C) \approx
\mathrm{Hom}(\C^{2^k},\C^{2^k})$, and $\C^{2^k}$ (the standard
fiber of the spinor bundle $\zeta$) is the simple module of
$M_{2^k}(\C)$, the $\mathrm{spin}^\C$ bundle has a cocycle
$\varphi_{\alpha\beta}(x) \in \mathrm{Aut}(\C^{2^k}) \subset
M_{2^k}(\C) \approx \mathrm{Cl}^{2k} $ of invertible elements $
\varphi_{\alpha\beta} \in \mathcal{F}^{2k}$ and
\begin{equation}\label{27}
    j(\varphi_{\alpha\beta}(x) )=\psi_{\alpha\beta}(x)\;.
\end{equation}

In general, not every manifold admits a $\mathrm{spin}^\C$
structure, and even if it admits there may exist different
$\mathrm{spin}^\C$ structures. In our example the base manifold
has only one (up to isomorphism) $\mathrm{spin}^\C$-structure.

The Dirac operator
\begin{equation}\label{28}
    D\; :\; C^\infty(\zeta) \; \longrightarrow \; C^\infty(\zeta)
\end{equation}
is determined by a linear connection on the spinor bundle and the
requirement that its symbol $\sigma(D)\; :\; T^\ast (E)
\longrightarrow \zeta^\ast \otimes \zeta $ be the unique
irreducible representation of the Clifford algebra
$\mathrm{Cl}^{2k}$ at each fibre (see \cite{Pal}).

In our example we choose $E$ to be a total space of a principal
$\mathrm{SU}(3)$ bundle over $M$ and $M$ to be isomorphic to
$\R^4$ as a topological manifold. We also choose the metric on $M$
to be $g_{\mu\nu}=\mathrm{diag}(-1,1,1,1)$, the basis
$\{\mathbf{f_\mu(x)}\}$ to be a non holonomic basis in $T(M)$, and
the metric $g_{0\;kl} (x) = g_{0\;kl}=\mathrm{diag}(-1,...,-1)$ to
be the Killing metric on the Lie algebra $su(3)$. We consider the
simplest case in which the ``scalar fields'' in Kaluza-Klein
ansatz are constants.
\\
\quad Because of (\ref{11}) the direct sum $T(E)=T^h(E) \oplus T^v
(E)$ is orthogonal. And from (\ref{17})
\begin{equation}\label{29}
    \mathrm{Cl}(TE)=\mathrm{Cl}(T^hE) \otimes \mathrm{Cl}(T^vE) \, ,
\end{equation}
   i.e., the standard fiber of $\mathrm{Cl}(TE)$ is $\mathrm{Cl}^{1,3} \otimes \mathrm{Cl}^{0,8}$ in the real (Majorana)
case and $\C^4 \otimes \C^8$ in the complexified case. So the
standard fiber of the spinor bundle $\zeta$ is isomorphic to $\R^4
\otimes \R^{16}$ in the Majorana case and $\C^4 \otimes \C^{16}$
in the complexified case. In the nonholonomic basis (\ref{7}) of
$TE$ according to (\ref{19}) we have
\begin{equation}\label{30}
    \begin{array}{ll}
      \mathbf{f}_\mu (x) \; \longmapsto & \; \gamma_\mu \otimes \mathbf{1} \in   \mathrm{Cl}(T^h_{(x,z)}E \oplus
 T^v_{(x,z)}E)=
   \mathrm{Cl}(T^h_{(x,z)}E,g) \otimes \mathrm{Cl}(T^h_{(x,z)}E,-g_0) \\
      \mathbf{e}_k (x) \; \longmapsto & \; \omega \otimes \gamma_{k} \in   \mathrm{Cl}(T^h_{(x,z)}E \oplus
 T^v_{(x,z)}E)=
   \mathrm{Cl}(T^h_{(x,z)}E,g) \otimes \mathrm{Cl}(T^h_{(x,z)}E,-g_0)  .
    \end{array}
\end{equation}
In the real case the Levi-Civita connection determines a unique
connection on the Majorana spinor bundle. Let $\nabla$ be the
Levi-Civita connection for the metric $g_{_E}$ on $E$ and
$\mathbf{h}_\alpha$ be an orthonormal with respect of $g_{_E}$
nonholonomic basis of $TE$. If
\begin{equation}\label{31}
    \nabla _{\mathbf{h_\alpha}}(\mathbf{h}_\beta)\equiv  \nabla _\alpha(\mathbf{h}_\beta)=
\Gamma_{\alpha\beta}^{E\,\rho}\,\mathbf{h}_\rho \, ,
\end{equation}
then in the real case there is a unique corresponding connection
on $\zeta$,
\begin{equation}\label{32}
    \begin{array}{l}
      \nabla_{\mathbf{h}_\alpha}=\mathbf{h}_\alpha +S_\alpha \, , \\
      S_\alpha =-\frac{1}{4}\Gamma_\alpha^{E\,\rho\sigma}\, \gamma_\rho \gamma_\sigma=
-\frac{1}{4} \Gamma _{\alpha\rho\sigma} \, \gamma^\rho \gamma^\sigma  \, ,\\
\Gamma_{\alpha\beta\sigma}=g_{\sigma\rho}^E
\Gamma_{\alpha\beta}^{\;\rho}\;,\ \; \quad
g_{\alpha\beta}^E= g^E (\mathbf{h}_\alpha,\mathbf{h}_\beta)=\mathrm{const}  \, . \\
    \end{array}
\end{equation}
If in the complexified  case we want to have a soldering between
the paralel transport in $TE$ and $\zeta$ there remains freedom
for the choice of the $U(1)$-connection. In our example we fix the
complexification of (\ref{32}). Then the Dirac operator reads:
\begin{equation}\label{33}
\begin{array}{l}
    D=\widehat{\mathbf{h}}^\mu \nabla_\mu=g^{E\,\mu\beta}\widehat{\mathbf{h}}_\beta\nabla_{\mu}=
\gamma^{E\,\mu}(\mathbf{h}_\mu+S_\mu)\ \ ,\quad
\ \ \widehat{\mathbf{h}}^\mu=\vartheta(\mathbf{h}^\mu) \\
    (D \psi)^a=\gamma ^{E\, \mu}(\mathbf{h}_\mu(\psi^a)+S_{\mu b}^{\ \,  a}\, \psi^b) \ \ , \
    \  \quad
\psi \in C^\infty(\zeta^\C) \, .  \\
\end{array}
\end{equation}
\quad For the Kaluza-Klein metric we specify
$g_{\mu\nu}=g(\mathbf{f}_\mu,\mathbf{f}_\nu)=\mathrm{const}=\mathrm{diag}(-1,1,1,1)
\ $, $ \ g_{0\;kl}=\mathrm{const}=\mathrm{diag}(-1,...,-1)$.
Because $\{\mathbf{f}_\mu\}$ is nonholonomic basis on $T(M)$,
"tetrada," $g$ can be an arbitrary metric on $M$. For simplicity
we take the "scalar fields" $g_{0\,kl}$ to be constant: the
Killing form in the Lie algebra $su(3)$. In the global basis
$\{\widehat{\mathbf{f}}_\mu, \mathbf{e}_k\}$ the metric (\ref{10})
has the form:
\begin{equation}\label{34}
\{g^E\}=
 \left(%
\begin{array}{cccccccc}
  -1 & 0 & 0 & 0 & 0 & . & . & 0 \\
  0 & 1 & 0 & 0 & 0 & . & . & 0 \\
  0 & 0 & 1 & 0 & 0 & . & . & 0 \\
  0 & 0 & 0 & 1 & 0 & . & . & 0 \\
  0 & 0 & 0 & 0 & -1 & 0 & 0 & 0 \\
  . & . & . & . & 0 & . & . & 0 \\
  . & . & . & . & 0 & . & . & 0 \\
  0 & 0 & 0 & 0 & 0 & . & . & -1 \\
\end{array}%
\right)\, .
\end{equation}
The coefficients of the Levi-Civita connection of the metric
(\ref{34})
\begin{equation}\label{35}
    \begin{array}{lllllllll}
      \nabla_{\widehat{\mathbf{f}}_\mu}(\widehat{\mathbf{f}}_\nu)
& \!\!\! = \!\! &
\Gamma^{E\;\rho}_{\mu\nu}\widehat{\mathbf{f}}_\rho+\Gamma^{E\;k}_{\mu\nu}\mathbf{e}_k
\; ; \quad
& \Gamma^E_{\mu\nu\rho} & \!\!\! = \!\! &
g_{\rho\sigma}\Gamma^{E\;\sigma}_{\mu\nu} \; , \quad
& \Gamma^E_{\mu\nu k} & \!\!\! = \!\! &
-g_{0\,ki}\Gamma_{\mu\nu}^{E \; i} \\
         \nabla_{\widehat{\mathbf{f}}_\mu}(\mathbf{e}_k)
& \!\!\! = \!\! & \Gamma^{E\;\nu}_{\mu
k}\widehat{\mathbf{f}}_\nu+\Gamma^{E\;l}_{\mu k}\mathbf{e}_l \; ;
& \Gamma^E_{\mu k \nu} & \!\!\! = \!\! &
g_{\nu\sigma}\Gamma^{E\;\sigma}_{\mu k} \; ,
& \Gamma^E_{\mu kl} & \!\!\! = \!\! &
-g_{0\,li}\Gamma_{\mu k}^{E \; i} \\
            \nabla_{\mathbf{e}_k}(\widehat{\mathbf{f}}_\mu)
& \!\!\! = \!\! &
\Gamma^{E\;\nu}_{k\mu}\widehat{\mathbf{f}}_\nu+\Gamma^{E\;l}_{k\mu}\mathbf{e}_l
\; ;
 & \Gamma^E_{k\mu\nu}
& \!\!\! = \!\! & g_{\nu\sigma}\Gamma^{E\;\sigma}_{k\mu} \; ,
& \Gamma^E_{k\mu l} & \!\!\! = \!\! &
-g_{0\,li}\Gamma_{k\mu}^{E \; i} \\
            \nabla_{\mathbf{e}_k}(\mathbf{e}_l)
& \!\!\! = \!\! &
\Gamma^{E\;\mu}_{kl}\widehat{\mathbf{f}}_\mu+\Gamma^{E\;m}_{kl}\mathbf{e}_m
\; ;
 & \Gamma^E_{kl\mu}
& \!\!\! = \!\! & g_{\mu\sigma}\Gamma^{E\;\sigma}_{kl} \; ,
& \Gamma^E_{ klm} & \!\!\! = \!\! &
-g_{0\,mi}\Gamma_{kl}^{E \; i} \\
    \end{array}
\end{equation}
are defined in the general case in (\ref{16}). Because of the form
of the metric in (\ref{34}) the Clifford bundle $\mathrm{Cl}(TE)$
has global generators:
\begin{equation}\label{36}
    \begin{array}{l}
      \gamma_\mu^E=\gamma_\mu \otimes \mathbf{1}\; , \  \quad \mu=1,2,3,4 \, , \\
      \gamma_k^E=\omega \otimes \gamma _k\; , \ \quad k=5,...,12 \, , \\
    \end{array}
\end{equation}
where $\gamma_\mu$ are generators of $\mathrm{Cl}^{1,3}$,
$\gamma_k$ are generators of $\mathrm{Cl}^{0,8}$ (\ref{30}),
$\omega=\gamma_1\gamma_2\gamma_3\gamma_4 $ $ \omega^2=-1$. After
these specifications the Dirac operator (\ref{32}), (\ref{36}) for
the Kaluza-Klein metric (\ref{10}) reads:
\begin{equation}\label{37}
    \begin{array}{lll}
      D & \!\! = \!\! &
\gamma
^{E\,\mu}\nabla_{\widehat{\mathbf{f}}_\mu}+\gamma^{E\,k}\nabla_{\mathbf{e}_k}=
(\gamma^\mu \otimes \mathbf{1})(\widehat{\mathbf{f}}_\mu +S_\mu)+
(\omega \otimes \gamma^k)(\mathbf{e}_k +S_k)  \vspace{3pt} \\
        & \!\! = \!\! &
(\gamma^\mu \otimes
\mathbf{1})\,(\mathbf{f}_\mu-A_\mu^{\;k}\mathbf{e}_k)
+ \ (\omega \otimes \gamma^k)\mathbf{e}_k \vspace{3pt} \\
        & \!\!  \!\! &
+ \  (\gamma^\mu \otimes \mathbf{1}) \lb -\frac{1}{4} \rb
(\Gamma_{\mu\nu\rho}^E\,\gamma^\nu \gamma ^\rho \otimes\mathbf{1}+
\Gamma_{\mu\nu k}^E\,(\gamma^\nu \otimes \mathbf{1})\,(\omega\otimes \gamma^k) \vspace{3pt} \\
        & \!\!   \!\! &
+ \ \Gamma^E_{\mu k \nu} (\omega \otimes \gamma^k)\,(\gamma^\nu
\otimes \mathbf{1}) +
  \Gamma_{\mu k l}^E\,(\omega \otimes \gamma^k)\,(\omega \otimes \gamma ^l))  \vspace{3pt} \\
        & \!\!  \!\! &
+ \ (\omega \otimes \gamma ^k )\lb -\frac{1}{4} \rb
(\Gamma_{k\mu\nu}^E \gamma^\mu \gamma ^\nu \otimes \mathbf{1} +
\Gamma_{k\mu l}^E (\gamma^\mu \otimes \mathbf{1}) \, (\omega \otimes \gamma^l) \vspace{3pt} \\
        & \!\!  \!\! &
+ \ \Gamma_{kl\mu}^E\,(\omega \otimes \gamma^l) \, (\gamma^\mu
\otimes \mathbf{1}) + \Gamma_{klm}^E\,(\omega \otimes \gamma^l) \,
(\omega \otimes \gamma^m)) \, . \\
    \end{array}
\end{equation}
Here $\gamma^\mu=g^{\mu\sigma}\gamma_\sigma$ and
$\gamma^k=-g_0^{ki}\gamma_i$.
 Using (\ref{16}) we obtain
\begin{eqnarray}\label{38}
 D&=&(\gamma^\mu \otimes \mathbf{1}) (\mathbf{f}_\mu-
 A_\mu^{\;k}\mathbf{e}_k) +( \omega \otimes \gamma^k) \mathbf{e}_k
 \nonumber \\
& &
 -\ \frac{\textstyle 1}{\textstyle 4}\, (\gamma^\mu \otimes
\mathbf{1}) ( \Gamma_{\mu\nu\rho}\,\gamma^\nu \gamma ^\rho
\otimes\mathbf{1}+ F_{\mu\nu k}\,\omega \gamma^\nu \otimes
\gamma^k  \nonumber \\
& & -\frac{\textstyle 1}{\textstyle 2}\,
A_\mu^{\;m}(t_{mlk}-t_{mkl}) \mathbf{1} \otimes \gamma^k\gamma^l )
- \frac{\textstyle 1}{\textstyle 4}\, ( \omega \otimes \gamma^k)
(\frac{\textstyle 1}{\textstyle 2}\, F_{\mu \nu k}
\gamma^\mu \gamma^\nu \otimes \mathbf{1} \nonumber \\
& & -A_\mu^{\;m}(t_{mlk}+t_{mkl})\omega  \gamma^\mu \otimes
\gamma^l- \frac{\textstyle 1}{\textstyle 2}
(t_{klm}+t_{mlk}+t_{mkl}) \mathbf{1} \otimes \gamma^l \gamma^m )\,
.
\end{eqnarray}
The Dirac operator (\ref{38}) for the Kaluza-Klein metric
(\ref{10}) acts on spinor fields which have 64 components. Due to
(\ref{30}) the standard fiber of the complex spinor bundle is
 $\C^4\otimes\C^{16}$.
\Section{Dimensional reduction of the Dirac operator} We need to
specify the action of the symmetry group $G=\mathrm{SU}(3)$ on the
spinor bundle. The action $R_g(x,z)=(x,zg)$ on the base $E=M\times
\mathrm{SU}(3)$ of the spinor bundle $\zeta$ must be lifted to a
bundle morphism action on the spinor bundle $\zeta$. This lifting
must be in agreement with the action of $\mathrm{SU}(3)$ on
$T(E)$. More precisely, let $R_\mathbf{g}:
(\mathbf{x},\mathbf{z})\,\rightarrow\,(\mathbf{x},\mathbf{z}\mathbf{g})$
be the action of $\mathbf{g}\in G=\mathrm{SU}(3)$ on $E$. For the
Kaluza-Klein metric (\ref{10}) the tangent lifting
 $R_\mathbf{g}^T:T_{(\mathbf{x,z})}(E)\,\rightarrow\,T_{(\mathbf{x,zg})}(E)$ is an isometry.
In our trivialization, in the basis $\{\widehat{\mathbf{f}}_\mu,
\mathbf{e}_k\}$ the tangent lifting
$R_\mathbf{g}^T:\R^{12}\,\rightarrow \,\R^{12}$ is the identity.
Let $\mathcal{F}\, : \C^4 \otimes \C^{16} \, \rightarrow \, \C^4
\otimes \C^{16}$ be the lifting of the action $L_\mathbf{g}$ to
the complexified spinor bundle $\zeta^{\C}$.
$\mathcal{F}_\mathbf{g}$ must satisfy (in the same trivialization)
\begin{equation}\label{39}
    j^\C(\mathcal{F}_\mathbf{g})=R^T_\mathbf{g} = \mathbf{1}
\end{equation}
with $j^\C$ given from (\ref{25}). Our purpose is to construct
explicitly the simplest example of the Dirac operator with
$\mathrm{SU}(3)$ symmetry and its reduction to the Dirac operator
acting on spinors over four-dimensional manifold. So we fix
$\mathcal{F}_\mathbf{g}=\mathbf{1}$ and then the action of
$G=\mathrm{SU}(3)$ on spinor fields, i.e., the sections on
$\zeta$, is
\begin{equation}\label{40}
    R_\mathbf{g} (\psi)^{\mu a} (\mathbf{x}, \mathbf{z})=\psi^{\mu a}(\mathbf{x}, \mathbf{z}\mathbf{g}) \, .
\end{equation}
\quad The Dirac operator (\ref{38}) for the Kaluza-Klein metric
(\ref{10}) is $\mathrm{SU}(3)$-invariant, when the action of
$\mathrm{SU}(3)$ on spinor fields is specified as in (\ref{40}).
For the invariant spinor fields, from (\ref{40}) we have
\begin{equation}\label{41}
    R_\mathbf{g} (\psi)^{\mu a}=\psi^{\mu a} \ \ \ \Rightarrow \ \ \
    \psi ^{\mu a} (\mathbf{x}, \mathbf{z}\mathbf{g})=
      \psi^{\mu a} (\mathbf{x}, \mathbf{e})\equiv \psi^{\mu a } (\mathbf{x}) \;.
\end{equation}
The set of all invariant spinor fields $C^\infty (\zeta)_G$ is
identified, due to (\ref{41}), with $C^\infty
(\zeta\mid_{M\times\{e\}}) \equiv C^\infty (\zeta \mid_{_M})$. The
dimensional reduction of the Dirac operator (\ref{38}) is a
restriction of (\ref{38}) on the set of $\mathrm{SU}(3)$-invariant
spinor fields and we obtain the reduced Dirac operator $D_r$:
\begin{equation}\label{42}
    D_r \,: \, C^\infty (\zeta \mid_{_M}) \; \rightarrow \; C^\infty(\zeta \mid_{_M}) \, .
\end{equation}
To calculate $D_r$ we have to put in (\ref{38}) a $\mathrm{SU}(3)$
invariant spinor field. For invariant spinor fields $\psi^{\mu
a}(\mathbf{x})$ we have from (\ref{41}), $\mathbf{e}_k(\psi^{\mu
a})=0$. For the reduced Dirac operator $D_r$ we obtain
\begin{equation}\label{43}
    D_r =(\gamma^\mu \otimes \mathbf{1})(\mathbf{f}_\mu-
        \frac{1}{4} \Gamma_{\mu\nu\rho}(x)\gamma^\nu\gamma^\rho) -
        \frac{1}{8}F_{\mu\nu k}(x) \omega \gamma^\mu \gamma^\nu \otimes
        \gamma^k+ \nonumber
        +\frac{1}{4}A_\mu^{\;m}(x) t_{mlk} \gamma^\mu \otimes \gamma^k \gamma^l +
        \frac{1}{4}t_{mlk}\omega \otimes \gamma^k \gamma^l \gamma^m ,
\end{equation}
where the coefficients $t_{mlk}$ are totally antisymmetric in all
indices.
 The reduced Dirac operator (\ref{43}) acts on the sections $\psi \in C^\infty (\zeta \mid_{_M})$.
The standard fiber is $\C^4 \otimes \C^{16}$. The bundle $\zeta
\mid_{_M}$ is canonically isomorphic to $\zeta^M \otimes
\zeta^{\mathrm{SU}(3)}$:
\begin{equation}\label{44}
    \zeta \mid_{_M} \approx \zeta^M \otimes \zeta^{\mathrm{SU}(3)} \, ,
\end{equation}
where $\zeta^M$ is the (complex) spinor bundle on $M$ and
$\zeta^{\mathrm{SU}(3)}$ is a vector bundle on $M$ with standard
fiber $\C^{16}$ considered as a simple module of the Clifford
algebra $\mathrm{Cl}^{0,8}$ corresponding to the Lie algebra
$su(3)$ with the Killing metric $g$. $\{\mathbf{f}_\mu\}$ is a
nonholonomic global basis of $T(M)$ and tetrada for the metric
$g$. $\Gamma_{\mu\nu\rho} (x)$ are the components of the
Levi-Civita connection of $g$ in the basis $\{\mathbf{f}_\mu\}$
and $-\frac{1}{4}\Gamma_{\mu\nu\rho} (x) \gamma^\nu \gamma^\rho$
are the components of the spinor connection in $\zeta^M$, so
\begin{equation}\label{45}
    D^M=(\gamma^\mu \otimes \mathbf{1})(\mathbf{f}_\mu -
         \frac{1}{4} \Gamma_{\mu\nu\rho}(x) \gamma^\nu\gamma^\rho)
\end{equation}
is the Dirac operator for the metric $g$ acting on spinor fields
with isotopic indices. $A_\mu^{\;m}(x)$ is a gauge field with
values in the Lie algebra $su(3)$ and $F_{\mu\nu k}(x)$ is its
curvature tensor. We can write (\ref{45}) in the form
\begin{equation}\label{46}
    D_r=(\gamma^\mu \otimes \mathbf{1})(\mathbf{f}_\mu -
         \frac{1}{4} \Gamma_{\mu\nu\rho}\gamma^\nu\gamma^\rho +
         \frac{1}{4} A_\mu^{\;m}t_{mkl}\mathbf{1} \otimes \gamma^k \gamma^l)
         -\frac{1}{8} F_{\mu\nu k} \omega \gamma^\mu \gamma^\nu \otimes \gamma^k +
         \omega \otimes \epsilon \, ,
\end{equation}
where $\epsilon=\frac{1}{4}t_{klm} \gamma^k \gamma^l \gamma^m$.

The interpretation of (\ref{46}) is that the reduced free massless
Dirac operator for the Kaluza-Klein metric acting on spinor fields
with an "isotopic" index on $M$ is equivalent to the usual Dirac
operator in the presence of a gravitation field (the metric) and
external gauge field with gauge group $\mathrm{SU}(3)$, source
term depending on the curvature $F_{\mu\nu}^{\; k}$ and a mass
term $\omega \otimes \epsilon$.

The "isotopic" bundle $\zeta^{\mathrm{SU}(3)}$ in (\ref{44}) has a
standard fiber $\C^{16}$. $\C^{16}$ is the unique simple module of
the Clifford algebra
$\mathrm{Cl}(su(3),-g_0)=\mathrm{Cl}^{0,8}=M_{16}(\R)$. The
algebra $su(3)$, as a real vector space, is isomorphic to $\R^8$
and $g_0$ is the negative defined Killing metric. In the chosen
basis $\{\widehat{\mathbf{e}}_k\}$ of $su(3)$,
$g_0=\mathrm{diag}(-1,...,-1)$. The group of symmetry of the
Killing metric $g_0$ on $su(3)$ (considered as a vector space,
isomorphic to $\R^8$) is $\mathrm{O}(8)$. According to the
standard procedure (\cite{ABS}, \cite{LM}), the Lie algebra $o(8)$
has a complex spinor representation in $\C^{16}$, which is a
direct sum of two irreducible representations and $\C^{16}=\C^8
\oplus \C^8$. These representations are realized on the
eigenspaces of the operator $\omega_0= \mathop {\prod}
\limits_{k=4}^{12} \gamma^k$. Let $\rho\; : \; \mathrm{O}(8)\;
\rightarrow \; \mathrm{End}(\C^{16})$ be the spinor
representation, $s \in o(8)$ and
$s(\widehat{\mathbf{e}}_i)=s_i^{\;j} \widehat{\mathbf{e}}_j$. Then
\begin{equation}\label{47}
    \rho(s)=-\frac{1}{4} s_{ij} \gamma^i \gamma^j \, .
\end{equation}
But the Lie algebra $su(3)$ has a natural adjoint representation:
$b \in su(3), ad(b) \in \mathrm{End}(su(3) \approx \R^6)$,
\begin{equation}\label{48}
\mathrm{ad}(\widehat{\mathbf{e}}_k)(\widehat{\mathbf{e}}_i)=
[\widehat{\mathbf{e}}_k,\widehat{\mathbf{e}}_i ] =
t_{ki}^{\;j}\widehat{\mathbf{e}}_j \, .
\end{equation}
The adjoint representation $\mathrm{ad}$ takes values in the Lie
algebra $so(8)$, i.e. $\{t_k\}\in so(8)$. So we can take the
composition of the two natural representations:
\begin{equation}\label{49}
    \begin{array}{l}
      \rho \circ \mathrm{ad} \; : \; su(3) \; \rightarrow \; \mathrm{End}(\C^{16}) \\
      (\rho \circ \mathrm{ad})(\widehat{\mathbf{e}}_k) =-\frac{1}{4} t_{kij} \gamma^i \gamma^j \, . \\
    \end{array}
\end{equation}
This is a representation of $su(3)$ in $\C^{16}$ which is a direct
sum of two eight-dimensional representations of $su(3)$. So the
bundle $\zeta^{su(3)}$ on $M$ is a Whitney sum of two
eight-dimensional bundles. Due to this fact, the $\mathrm{SU}(3)$
invariant spinors, i.e., the sections $C^\infty(\zeta
\mid_{_M})=C^\infty(\zeta^M \otimes \zeta^{\mathrm{SU}(3)})$ have
the natural interpretation as two $su(3)$ spinor octets.

The main result of this paper is that the free massless
$\mathrm{SU}(3)$-invariant Dirac operator on the manifold $E$ (the
total space of principal $\mathrm{SU}(3)$-bundle on
four-dimensional manifold $M$) after dimensional reduction is
equivalent to the Dirac operator on four-dimensional manifold $M$,
acting on two $\mathrm{SU}(3)$ spinor octets, in the presence of
gravitational field, external $\mathrm{SU}(3)$ gauge field with a
source depending on the curvature tensor of the $\mathrm{SU}(3)$
gauge field and mass term as it is in (\ref{46}).
\Section{Comments} One of the ideas of the "Kaluza-Klein approach"
is that a collection of fields with different nature involved in
complicated differential equations may be considered as ``simple''
differential equations for one-type field in a multidimensional
case, but having some symmetry, and considered only on the
invariant fields. In the spirit of this idea we comment here on
the steps in the reduction procedure, where the structures of new
type arise. In this example $E$ is 12-dimensional manifold with
trivial tangent bundle. For arbitrary metric $g_{_E}$ on $E$ there
is just one spinor structure and the spinor fields have
64-components. The group of symmetry $G=\mathrm{SU}(3)$ acts on
$E$ as on a total space of a principal bundle. This separate the
vertical subbundle $T^v(E)\hookrightarrow T(E)$. The metric under
consideration $g_{_E}$ determines a horizontal subbundle
$T^h(E)\hookrightarrow T(E)$ as an orthogonal complement of
$T^v(E)$. The metric $g_{_E}$ is $G$-invariant, so the horizontal
subbundle $T^h(E)$ is $G$-invariant and is a linear connection
with structure group $G=\mathrm{SU}(3)$. Further, the linear
connection in the spinor bundle, coming from the Levi-Civita
connection for $g_{_E}$ and needed for the Dirac operator, is
expressed in terms of this $\mathrm{SU}(3)$-connection. This leads
to the appearance of the $\mathrm{SU}(3)$ gauge field and its
stress tensor. These are classical results for the Kaluza-Klein
ansatz. The orthogonal splitting $T(E)=T^h(E)\oplus T^v(E)$
according to the classifying theorem for Clifford algebras leads
to the representation (\ref{29}) and to appearance of spinors on
four-dimensional base manifold after the reduction. The metric in
the vertical subspace is the Killing metric. There is a natural
adjoint representation of Lie algebra $su(3)$ on itself,
orthogonal to the Killing metric and a natural representation of
the orthogonal group of the Killing metric on corresponding
16-dimensional spinors. Due to the representations (\ref{29}),
this leads to the appearance of two $\mathrm{SU}(3)$ spinor octets
after the reduction. Finally, the $G$-invariance of the spinor
fields in the simple case that we consider leads to vanishing of
the vertical derivatives and we obtain the reduced operator acting
on the fields defined on the four-dimensional base manifold.

\end{document}